\documentclass[10pt,conference,compsocconf]{IEEEtran}
\usepackage{times}
\usepackage[T1]{fontenc}
\usepackage[latin9]{inputenc}
\usepackage{booktabs}
\usepackage{mathtools}
\usepackage{textcomp}
\usepackage[inline]{enumitem}
\usepackage{amsmath,amsthm,amssymb}
\usepackage{caption}
\usepackage{graphicx}
\usepackage{array}
\usepackage{mdwmath}
\usepackage{mdwtab}
\usepackage{eqparbox}
\usepackage{fixltx2e}
\usepackage{stfloats}
\usepackage{siunitx}
\usepackage[linesnumbered,ruled,vlined]{algorithm2e}
\usepackage{url}
\providecommand{\tabularnewline}{\\}
\begin{document}
\bstctlcite{IEEEexample:BSTcontrol}
\pagenumbering{gobble}

\title{\textbf{\Large A Reliable IoT-Based Embedded Health Care System for Diabetic Patients}\\[0.2ex]}

\author
{\IEEEauthorblockN{Zeyad A. Al-Odat\IEEEauthorrefmark{1},
Sudarshan K. Srinivasan\IEEEauthorrefmark{1}, 
Eman M. Al-Qtiemat\IEEEauthorrefmark{1}, 
Sana Shuja\IEEEauthorrefmark{2}
}
\IEEEauthorblockA{\IEEEauthorrefmark{1}Electrical and Computer Engineering, 
North Dakota State University\\
Fargo, ND, USA\\
\IEEEauthorrefmark{2}Electrical Engineering, COMSATS Institute of Information Technology, \\
Islambad, Pakistan\\
Emails: \IEEEauthorrefmark{1}zeyad.alodat@ndsu.edu,
\IEEEauthorrefmark{1}sudarshan.srinivasan@ndsu.edu,
\IEEEauthorrefmark{1}eman.alqtiemat@ndsu.edu,\\
\IEEEauthorrefmark{2}SanaShuja@comsats.edu.pk
}
}
\maketitle

\begin{abstract}
This paper introduces a reliable health care system for diabetic patients based on the Internet of Things technology. A diabetic health care system with a hardware implementation is presented. The proposed work employs Alaris 8100 infusion pump, Keil LPC-1768 board, and IoT-cloud to monitor the diabetic patients. The security of diabetic data over the cloud and the communication channel between health care system components are considered as part of the main contributions of this work. Moreover, an easy way to control and monitor the diabetic insulin pump is implemented. The \mbox{patient\textquotesingle s} records are stored in the cloud using the Keil board that is connected to the infusion pump. The reliability of the proposed scheme is accomplished by testing the system for five performance characteristics (availability, confidentiality, integrity, authentication, and authorization). The Kiel board is embedded with Ethernet port and Cortex-M3 micro-controller that controls the insulin infusion pump. The secure hash algorithm and secure socket shell are employed to achieve the reliability components of the proposed scheme. The results show that the proposed design is reliable, secure and authentic according to different test experiments and a case study of the Markov model. Moreover, a 99.3\% availability probability has been achieved after analyzing the case study.    

\end{abstract}


\IEEEpeerreviewmaketitle

\section{Introduction}
Cloud computing has been integrated with the Internet of Things (IoT) to enable the network devices to provide resilient services to all users and applications over the world. This integration helps to simplify the access of the IoT-enabled devices by all kind of users and applications, e.g., physical devices~\cite{alodat2018iot}. IoT is able to connect ubiquitous systems (including physical devices) using different network infrastructures to provide efficient services all the time~\cite{lin2017survey}.  

The physical devices that are linked to the (IoT) are continuously increasing and emerging, which put a burden on the IoT service providers to provide secure and efficient services~\cite{al2015internet}. Physical devices are allowed to mimic human being's senses through various software and hardware that are connected together using the IoT. For example, the use of a smart home as an IoT-based application can turn on and off the air conditioning system when sensing the home residents leaving or coming their home~\cite{atzori2010internet}. Moreover, IoT-enabled devices can be controlled using a web page or smartphone applications, in the presence of Internet~\cite{kazi2015iot}. 

To utilize the IoT more efficiently, the industrial world has moved toward the use of IoT in small board and chips. For instance, manufacturers enable the internet connection on their small boards by adding the internet accessibility option to their products~\cite{ungurean2014iot}. Moreover, different primitives can be connected together through IoT-based applications, and they can access a shared medium between them in the presence of IoT-cloud, e.g., the health care records that are shared between the patient, hospital, and eligible users can be accessed over the cloud through mobile applications~\cite{hinge2013mobile}.

The security and authenticity of the IoT-based applications become crucial, because many entities joined the world of IoT, and the possibilities of attacks and collisions have increased~\cite{conti2018internet}. Therefore, the term of "Cyber-Physical System" (CPS) emerged to provide the integration between physical devices and cyber security~\cite{lin2017survey}. Particularly, the integration of the IoT-base health care records where the health records are saved on the cloud and shared with different entities. Moreover, recent improvements in the IoT designs help with the support of health care systems, e.g.,  the tracking \mbox{patient\textquotesingle s} records and bio-medical devices using the IoT applications~\cite{catarinucci2015iot}\cite{yu2016enabling}. 

Medical devices for diabetic care have also joined the world of IoT by supporting versatile design options~\cite{gai2015electronic}. However, security issues need to be addressed to ensure device security and the \mbox{patient\textquotesingle s} privacy~\cite{chen2015new}. A system with an authentic security mechanism is required to guarantee the integrity and security of \mbox{patient\textquotesingle s} records. One of the existing methods that can be easily implemented in hardware is the Secure Hash Algorithm (SHA)~\cite{harsha2014design}. The SHA is an official hash algorithm standard that was standardized by the National Institute of Standards and Technology (NIST)~\cite{dang2013changes}.

SHA is compatible with hardware-level implementation, which makes it the most desirable methods for hardware designers to implement their reliable architectures~\cite{pub2012secure}. The implementation of IoT technology in hardware has become crucial for high-performance applications~\cite{salinas2018efficient}. The hardware allows a high-speed computation to manipulate and retrieve health records where health records are increasing day after the other. Therefore, medical-hardware designers have moved toward the use of IoT hardware-units in their designs to support high-speed computation power for IoT related \mbox{functions~\cite{boppudi2014data}.} 

This paper introduces an IoT-based embedded scheme for a diabetic insulin pump. The proposed design elaborates the mechanisms of data acquisition and monitoring between different parties (patient, cloud, hospital, and legitimate users). This design helps to share health data that are related to a \mbox{patient\textquotesingle s} diabetes disease along with other health records on the cloud. All these data need to be secured and authenticated when they are retrieved from the cloud. 
We use the SHA algorithm to provide the security and authenticity terms for our proposal.  

The rest of the paper is organized as follows. 
Section~\ref{sec:section2} provides preliminaries about the used components in this paper. Section~\ref{sec:section3} presents a literature review about the related work. The proposed methodology is presented in Section~\ref{sec:section4}. Results and discussions are detailed in Section~\ref{sec:section5}. Section~\ref{sec:section6} concludes the paper.

\section{Preliminaries}
\label{sec:section2}
Before going through the details of our proposal, brief descriptions about SHA-256, health care system components, and performance characteristics are presented in the subsequent text.

\subsection{Brief Description of the SHA-256}
SHA-256 is employed in our design to provide data integrity and authenticity. SHA-256 takes a message with an arbitrary size then, through message compression operations, produces a message hash of size 256-bit. Equation~(\ref{eq:hash}) shows how to get the hash ($h$) from a message ($M$) using compression function ($H$).

\begin{equation}
h = H(M),
\label{eq:hash}
\end{equation}
where $M$ is the input message and $h$ is the digest generated using the hash algorithm $H$.

The secure hash algorithm is used to make sure that the data have not tampered during transmission. For instance, the message hash is computed at the sender side and appended with the transmitted message, then at the receiver side the received message hash is recomputed again and compared with the appended hash value. For the unchanged message, the hash values on both sides are equal, which means that the message has not tampered during the transmission.

\begin{figure*}[htbp]
    \centering
    \includegraphics[width=0.75\textwidth]{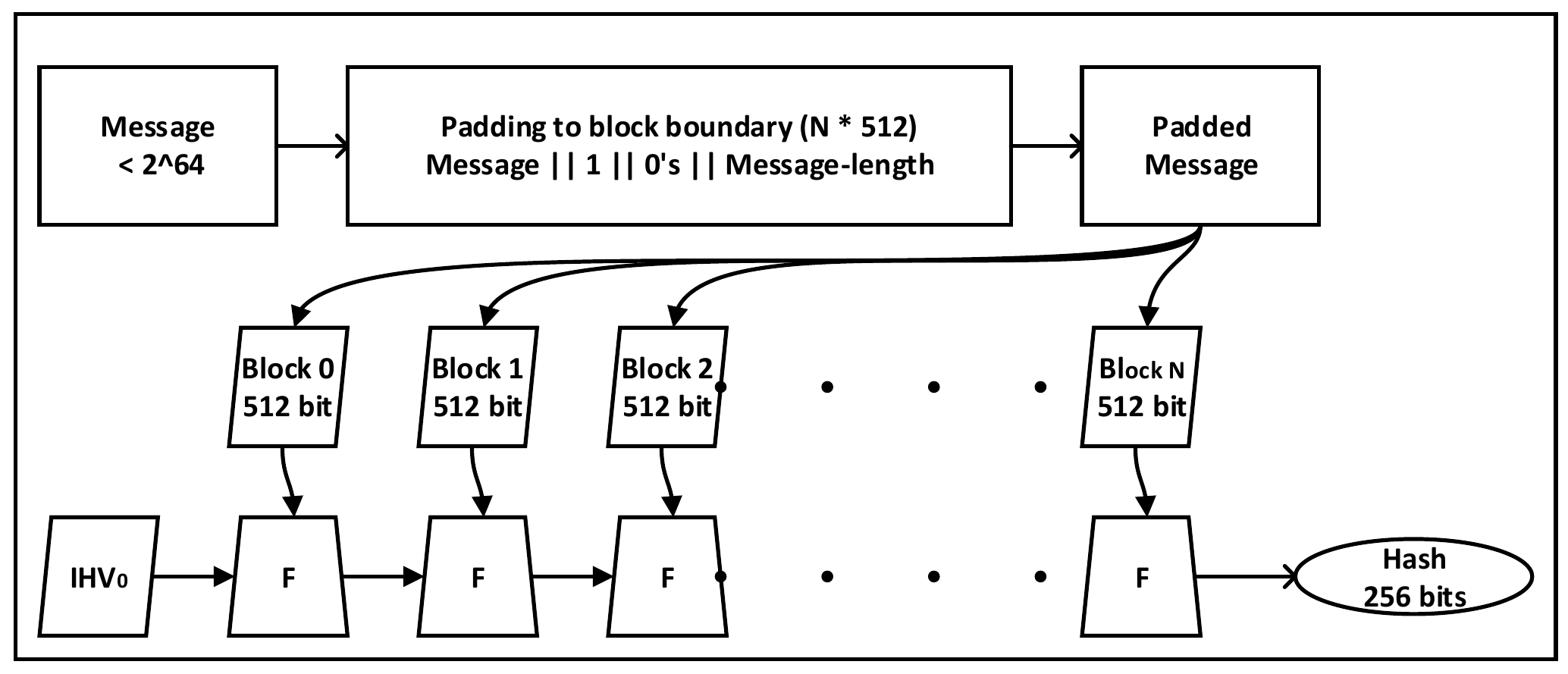}
    \caption{General architecture to compute the SHA-256 hash function.}
    \label{fig:sha}
\end{figure*}

Figure~\ref{fig:sha} depicts the general procedure that is used to compute the SHA-256 hash for any given message. The input message of size less than $2^{64}$ is padded first by adding $1$ at the end of the message then add the least number of zeros to make it congruent to $448/512$. then the message size is appended to the end of the message as a 64-bit. At the end of the pre-processing phase, the final message size becomes multiple of $512$-bit. Afterward, each message block is processed using the Initial Hash Value ($IHV_0$) and SHA-256 compression function ($F$). The output of each block is fed as $IHV$ to the next block calculations.  

At the end of the process, the hash value that is generated from the last block produces the final $256$ bits hash. A detailed description of the secure hash algorithm can be found in~\cite{pub2012secure}.

Unlike the secure hash algorithm, the keyed-hash message authentication code (HMAC) involves a secure hash algorithm and a secret cryptography key. But, the \textit(HMAC) algorithm is vulnerable against the length extension attack, which gives the attacker an opportunity to access the secret data~\cite{lengthattack}. Therefore, we avoid using the \textit{HMAC} algorithm in our design. Though, data encryption functionality is the responsibility of the employed hardware and the encrypted \textit{SSH} connection.  

\subsection{System Components}
The proposed design consists of components that integrate together to form the overall architecture.  
\begin{itemize}
    \item Micro-controller unit. It is used to manage and control the medical devices according to a predefined procedure. This includes: delivers the control commands, daily \mbox{patient\textquotesingle s} readings, and provide the secure connection layer. In our design, we use the Cortex-M LPC-1768 Keil board.  
    \item Infusion Pump. It delivers the medical liquid (insulin) to the patient on a timely basis. In our design, Alaris-8100 infusion pump module is used.
    \item IoT-based cloud storage. In our proposal, we use the IoT-cloud as a medium between distributed medical institutions, patients and caregivers. 
    \item Security components. They include a secure communication path using the secure socket layer (SSL/TLS), and cryptography mechanism to ensure the security of all system components.
    \item Legitimate users. The list of all authorized users to use the system according to predefined privileges.
\end{itemize}

\subsection{Performance Characteristics}
Today, some medical liquids are delivered programmatically without human intervention, e.g., insulin~\cite{paul2011review}.
With medical devices that include embedded systems, a number of conditions need to be met to consider them as reliable and secure systems.

\begin{itemize}
\item Availability. The property that gives the probability of the system being in the normal state for a period of time.
\item Confidentiality. The property that ensures the \mbox{patient\textquotesingle s} information and system data are unavailable to unauthorized third parties.
\item Integrity. All system data that can affect the treatment of the patient must not be altered without the \mbox{patient\textquotesingle s} knowledge.
\item Authentication. It means, only authorized parties or components should be able to act as a trusted user of the system.
\item Authorization. The property of providing the verification of certain actions before execution.
\end{itemize}
These characteristics will be discussed in Section~\ref{sec:section5}. 

\subsection{Contributions}
The proposed design aims to provide the following contributions to the health care system, particularly diabetic patients. We use an external micro-controller (Kiel LPC1768) to program Alaris-8100 infusion pump. This design helps to solve current problems in the infusion pump.    
\begin{itemize}
    \item On-time medication, where a patient can get all his prescribed doses on time. 
    \item Simplicity, affordability and the ease of use.  
    \item Remote health record management through mobile applications or web browsers.
    \item Provide health service on the time of Off-Service physician. 
    \item Provide secure and authentic health care service by employing cryptography and security approaches. 
\end{itemize}

\section{Related Work}
\label{sec:section3}

Recently, the IoT-based applications have involved in all fields that influence Human life, especially, medical devices. The use of IoT in health monitoring and control is employed by different publications~\cite{catarinucci2015iot}\cite{harsha2014design}\cite{rahmani2015smart}\cite{hsueh2016next}\cite{liu2015secure}. A novel IoT-aware smart architecture for automatic monitoring and tracking of the patient, personnel, and biomedical devices, was presented in \cite{catarinucci2015iot}. The proposed work built a smart hospital system relying on three components: Radio Frequency Identification (RFID), Wireless Sensor Network (WSN), and smart mobile. The three hardware components were incorporated together through a local network to collect the surrounding environment and all related parameters to a \mbox{patient\textquotesingle s} physiology. The collected data is sent to a control center in a real-time manner where all data are available for monitoring and management by the specialist through the Internet. The authors implemented a Graphical User Interface (GUI) to make the data access more flexible for the specialist.  

To exploit the bridging point between the IoT and health care system, Rahmani~\textit{et al.} proposed a smart E-health care system for ubiquitous health monitoring~\cite{rahmani2015smart}. The proposed work exploits ubiquitous health care gateways to provide a higher level of services. This work studied significant ever-growing demands that have an important influence on health care systems. The proposed work suggests an enhanced health care environment where control center burdens are transferred to the gateways by enabling these gateways to process part of the control center jobs. The security of this scheme was taken into consideration as the system deals with substantial health care data. The security scheme provides data authenticity and privacy characteristics.  

A personalized health care scheme for the next generation wellness technology was proposed in~\cite{hsueh2016next}. 
The security of \mbox{patient\textquotesingle s} records was addressed in case of data storage and retrieval over the cloud. 
The proposed work established a patient-based infrastructure allowing multiple service providers including the patient, service providers, specialists, and researchers to access the stored data. Their work was implemented on a cloud-based platform for testing and verification where a customized and timely messaging system for continuous feedback is tested. Moreover, multiple service providers are supported with an information infrastructure to provide unified views of \mbox{patient\textquotesingle s} records and data. 
The use of special encryption schemes was also explored in~\cite{liu2015secure,8412485}. Liu \textit{et al.} presented a scheme for secure sharing of personal health records in the cloud. The health records are ciphered before they are stored in the cloud. The proposed work uses Cipher-Text Attribute-Based Signcryption Scheme (CP-ABSC) as an access control mechanism. Using this scheme, they were able to get fine-grained data access over the cloud~\cite{liu2015secure}. While Zhang~\textit{et al.} proposed a cloud storage scheme for electronic health records based on secret sharing. The proposed design consists of four phases, namely, the preprocessing phase, distribution phase, reconstruction outsourcing phase, and recovery and verification phase. In the preprocessing phase, each health record is uploaded to the cloud as a set of $m$ blocks. Then in the distribution phase, the blocks are distributed over different storage locations in the cloud. In the reconstruction phase, the record's blocks are gathered from different storage locations. Lastly, in the verification phase, the gathered blocks are verified to determine whether if they belong to the accurate record or not~\cite{8412485}.

With the emerge of IoT-enabled micro-chips, the researchers got benefited from this property by implementing embedded systems that provide IoT capabilities~\cite{joyia2017internet}. Different publications explored the use of embedded micro-controllers in medical devices. Particularly, the use of Keil LPC1768 micro-controller~\cite{harsha2014design}\cite{boppudi2014data}. 
In~\cite{harsha2014design}, an online design for monitoring \mbox{patient\textquotesingle s} data was presented. The proposed work employed an Advanced RISC Machine (ARM) architecture where Cortex M3 microprocessor is embedded in Keil LPC1768 board. In their work, the authors used pulse, temperature, and gas sensors to collect the \mbox{patient\textquotesingle s} medical parameters. The LPC1768 board was used as a hardware layer between the Internet and the medical sensors. Each time the sensors' values change, the corresponding values on the Internet change immediately. However, their design was only used to monitor the surrounding environment without any interaction with the patient. 

To have an embedded system with monitoring and control capabilities, Boppudi \textit{et al.} proposed a data acquisition and control system using the ARM Cortex M3 microprocessor~\cite{boppudi2014data}. The proposed design send the monitored sensor data to the Internet using an Ethernet-controlled interface, which was built using Keil LPC1768 board. 
The proposed work employed two sensing devices temperature and accelerator-meter. Both sensors were used to collect data from the surrounding environment. The collected readings are sent to the Internet through the Ethernet interface. According to the uploaded readings, a specialist can change the behavior of the device through the Internet browser.

With the distributed components of the IoT-based health care systems, the need to verify and evaluate the integration of these components is crucial. The verification and evaluation of health care systems over the cloud is investigated by different researchers~\cite{macedo2014dependability,anastasiia2018markov}.  
Macedo \textit{et al.} proposed a model to evaluate the IoT-based data redundancy. They employed a Markov model to test the probability of failure of one of the IoT components during the run time. They calculated the probability of failure of one of the cloud storage, then transfer the data store burden to less probability storage devices. The proposed design investigates the failure probability of the cloud storage components using the failure and recovery factor of each component. They were able to build a Markov model that describes the transition between the redundant storage locations at any given time~\cite{macedo2014dependability}. However, Anastasiia \textit{et al.} extended their work to build a model for IoT health care system~\cite{anastasiia2018markov}. The proposed work establishes a Markov model considering the failure of components for the IoT health care system. In their work, they gave the case study of Markov model to test the availability of health care components if any failure has happened at any time or location.  

In the subsequent section, the integration between different components of the IoT health care system and the conjunction between the diabetic insulin pump (Alaris 8100) and Keil LPC-1768 board will be discussed in details. 

\section{Proposed Methodology}
\label{sec:section4}
In the proposed methodology, all system components that were mentioned in Section~\ref{sec:section2} will be integrated together to form the general architecture of the embedded IoT health care system. The proposed design comprises three main operations: monitoring, storing, and control, which are connected together to form the overall system. In this section, a case study of Markov model will be presented to test the availability of the proposed design. 

For secure communication, the Secure Socket Shell protocol is employed. The SSH is the worldwide highest quality level for remote framework organization and secure document exchange. SSH is utilized in each datum focus and in each real endeavor. One of the highlights behind the enormous prevalence of the SSH is the solid verification utilizing SSH keys~\cite{williams2011analysis}.
\subsection{General Architecture of the Proposed Scheme}
\begin{figure}[tbp]
    \centering
    \includegraphics[width=\columnwidth]{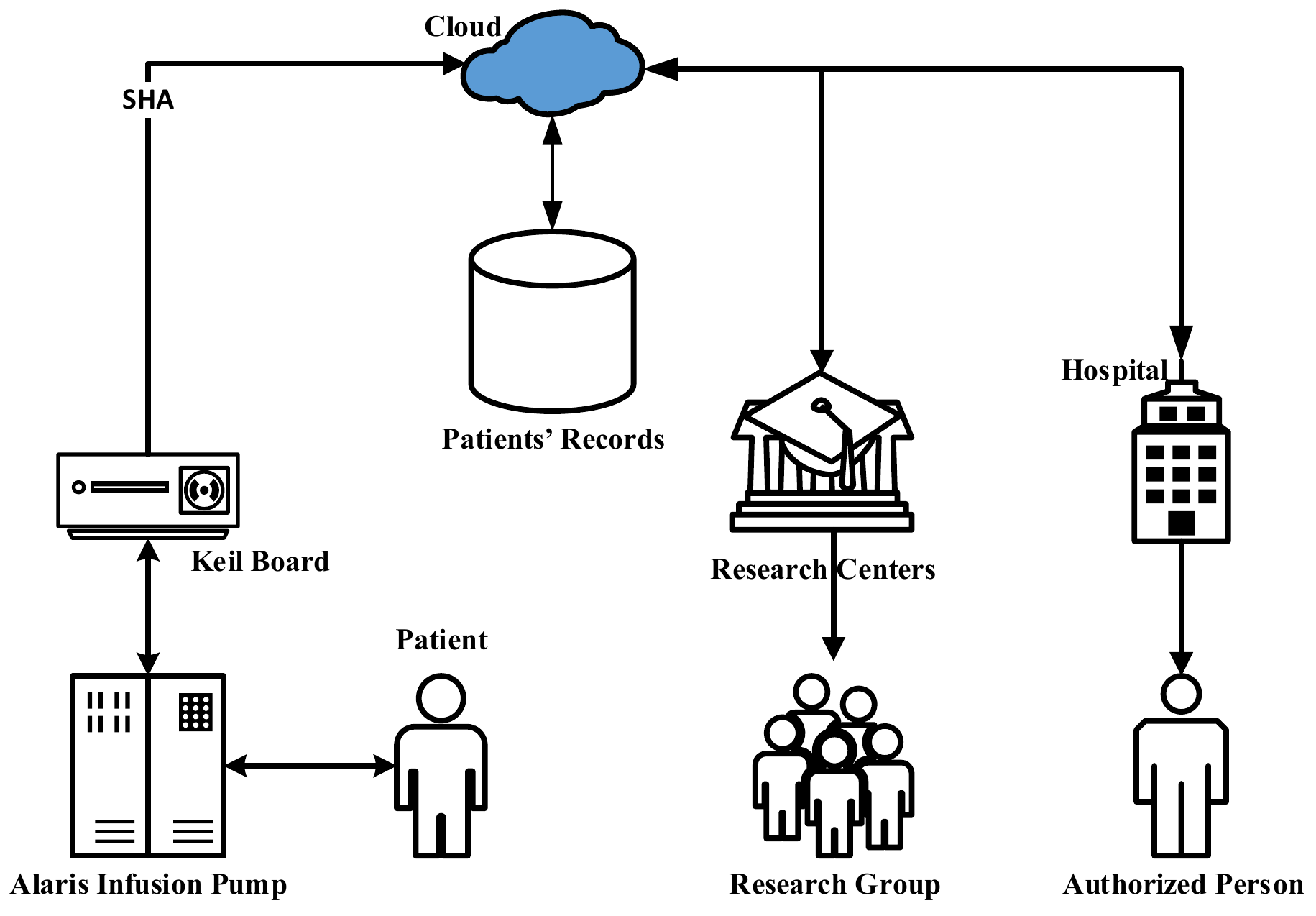}
    \caption{General architecture of the proposed scheme.}
    \label{fig:general}
\end{figure}

The proposed design employs the Alaris 8100 infusion pump to deliver insulin to the patient. The infusion pump is controlled using LPC-1768 board that contains the Cortex-M3 micro-processor. Figure~\ref{fig:general} shows the general architecture of the proposed design.   

The diabetic patient is attached to the infusion pump to get prescribed insulin doses. The Infusion pump is connected to the micro-controller unit (Keil LPC-1768 board) through a serial connection. A secure connection between the micro-controller and the cloud is established using the Secure Socket Shell (SSH) protocol and supported by the SHA-256 mechanism to authenticate the data exchange between cloud and micro-controller. Cloud computing provides the required infrastructure to handle all communications between the local and remote entities and reserves the desired amount of storage to store all health records and \mbox{patient\textquotesingle s} data. The proposed architecture allows the authorized remote entities (e.g., medical and research institutions) to access the stored health records and monitor the \mbox{patient\textquotesingle s} vital signs. Moreover, the proposed architecture provides the ability to control the infusion pump, remotely, through privileges that are given to an authorized physician. 

\begin{figure}[htbp]
    \centering
        \includegraphics[width=\columnwidth]{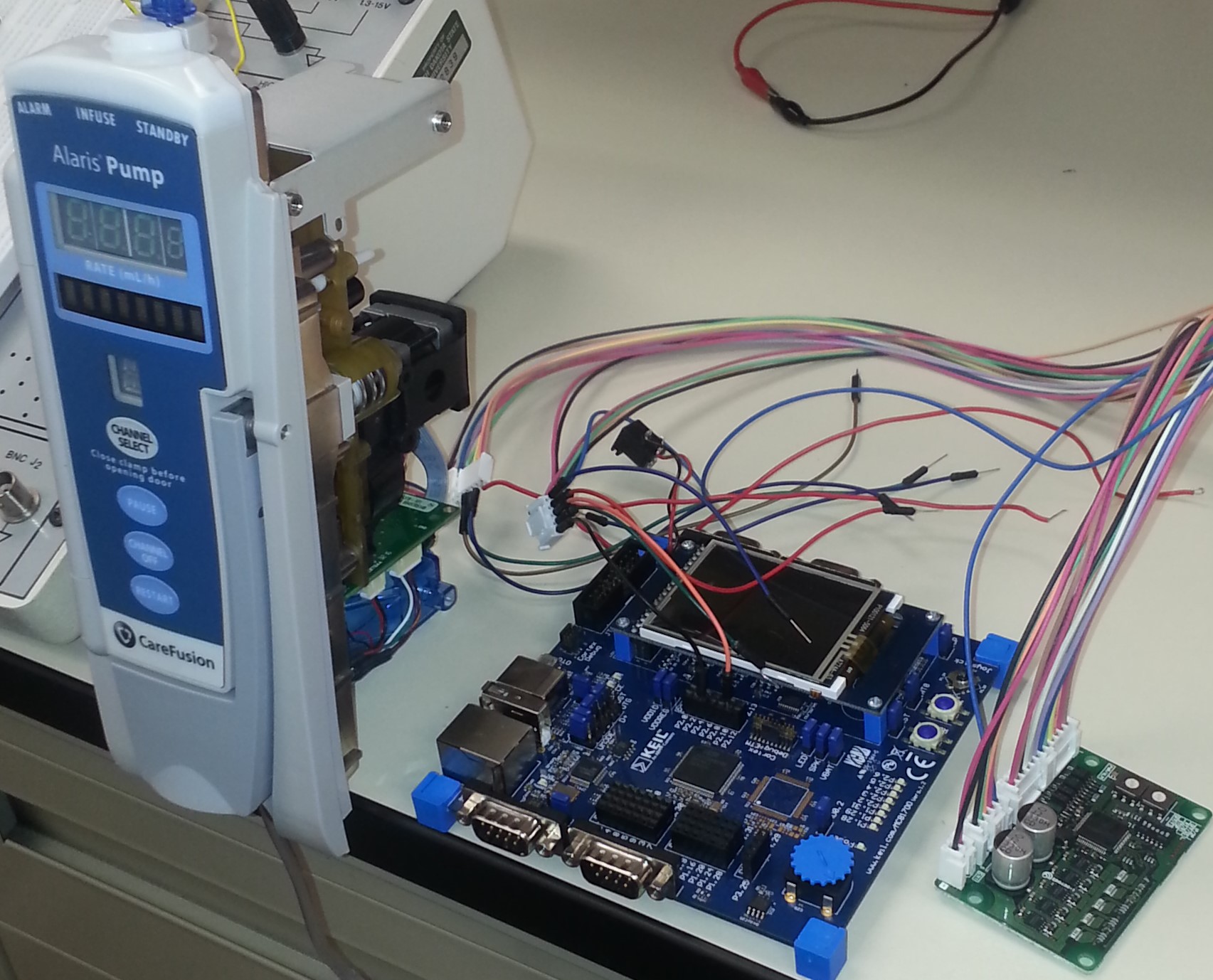}
    \caption{Connection of Alaris Infusion Pump 8100 with Keil 1768 PCB board.}
    \label{fig:Circuit}
\end{figure}

Figure~\ref{fig:Circuit} shows the hardware setup of the proposed architecture. The Alaris-8100 infusion pump was disassembled to reach out the infusion components inside the pump. Then we built the interface between the Keil LPC-1768 board and the pump. Afterward, we used Keil $\mu$-Vision Software Development Kit (SDK) to program the micro-controller. 

The hardware setup operations and system deployment were integrated together at the North Dakota State University (NDSU)-Electrical and Computer Engineering laboratories.

\subsection{Monitoring, storing and controlling IoT health care system}
The proposed design categorizes the IoT-health care system into three operations, which are the monitor, store, and control operations. The monitor operation involves the process of monitoring the status of the patient at any time and broadcasts the recorded data to the legitimate parties. The monitoring operation is accomplished by the micro-controller and insulin pump sensors. The store operation responsible for storing the collected data in local and remote databases, which is accomplished by the micro-controller. The control operation, which is accomplished by the micro-controller, changes the insulin pump schedule according to predefined or modified schedules. The schedule of the insulin pump is only generated by an authorized physician. Each operation is a complement to the other where the micro-controller operates as a common part between them.

\subsubsection{Monitor health records}
The process of health record monitor is accomplished according to Algorithm~\ref{algo:monitor}. The Secure Socket Shell (S) connection is initialized between the legitimate user and the cloud. Then the legitimate user receives the desired patient record appended with its SHA-256 hash value ($H_p$). The hash value ($H_q$) of the received record ($P_q$) is computed at the user side then, compared with the appended hash value ($H_p$). If both hash values are equal then the received health record is valid and contains the last updated health data. 

\begin{algorithm}[htbp]
	\DontPrintSemicolon 
	\KwIn{Query ($Q$)}
	\KwOut{$Q$ + Hash($c$)}
	\For{$q \gets 0$ \textbf{to} $n$} {
	    $S = Init(SSH)$\\
	    $Receive(P_q + H_p)$\\
		$H_q = Hash(P_q)$\\
		$Compare(H_q, H_p)$\\
		\quad \quad $Case(equal)$ $\gets$ \texttt{Valid}\;
		}
	\caption{Monitor \mbox{patient\textquotesingle s} records}
	\label{algo:monitor} 
\end{algorithm}

\subsubsection{Store health records}
Each health record has a designated SHA-256 value that is appended to the health record at the time of generation. Algorithm~\ref{algo:store} shows the general procedure that is carried out to store the newly generated or updated health record. The hash value ($H_p$) of health record ($P$) that is related to the patient ($i$) is computed using the SHA-256 hash function. The computed hash ($H_p$) is appended to the patient record ($P_i$). An SSH connection between the micro-controller and the cloud is initialized to send the combination of hash and record ($A_p$) to the cloud for storage. Moreover, the new health record is stored in a Local Storage (LS) unit for quick data access.      
\begin{algorithm}[]
	\DontPrintSemicolon 
	\KwIn{Health record ($P$)}
	\KwOut{$P$+Hash($P$)}
	\For{$i \gets 0$ \textbf{to} $n$} {
		$H_p = Hash(P_{i})$\\
		$A_p = Append(P_{i},H_p)$\\
		$S = Init(SSH)$\\
		$LS(A_p)$\\
		$Send(A_p,S)$\;
		}
	\caption{Store health records}
	\label{algo:store} 
\end{algorithm}

\begin{figure}
     \centering
     \includegraphics[width=0.75\columnwidth]{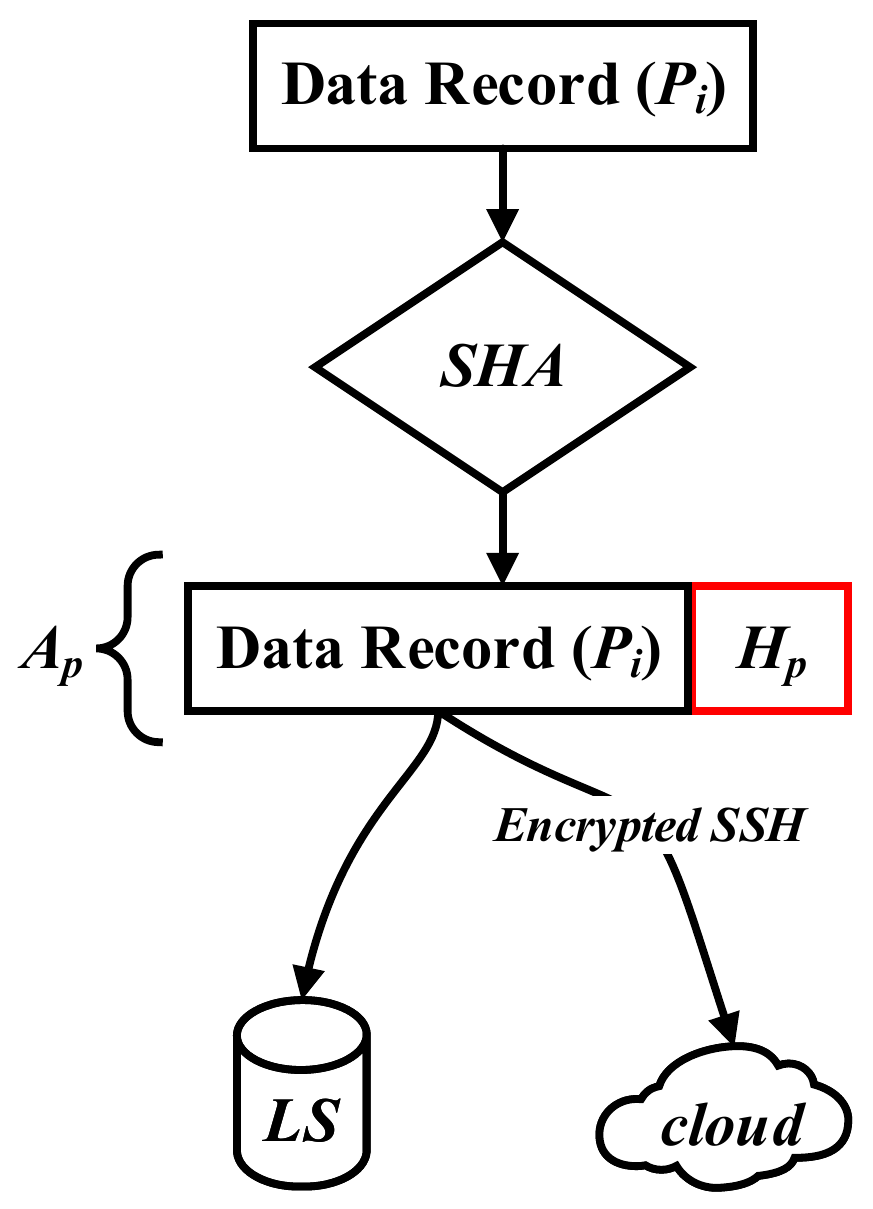}
     \caption{Construction of data record.}
     \label{fig:data_record_construction}
 \end{figure}

 As health records are sensitive information, the \textit{SSH} uses a symmetric encryption mechanism to ensure the data privacy between different parties. This is accomplished after initialization of the \textit{SSH} connection between client and server. The client initializes the connection by contacting the server, then the server responds to the client by sending the server's public key. Figure~\ref{fig:data_record_construction} shows the construction of data record ($P_i$). The data record is signed using the \textit{SHA} algorithm, then the produced hash value ($H_p$) is appended to the end of the data record. Afterward, The \textit{SSH} connection is used to transfer data record to the cloud. 
 
 {}
 
\subsubsection{Prescription control command}
The prescription control command is generated by a remote caregiver. Algorithm~\ref{algo:send} shows the general procedure to send a new control command to the insulin pump. The prescription control command ($C$) is generated and appended with its corresponded SHA-256 hash value ($H_c$) to form the appended control command ($A_c$). A secure Socket Shell ($S$) is initialized between the remote caregiver and the micro-controller through the cloud. Then the new control command is sent through the $SSH$ Chanel. At the receiving side, the micro-controller verify the received control command by following steps $3-6$ of Algorithm~\ref{algo:monitor} where the received message is $C+H_c$. If the received prescription control command is valid, then the micro-controller will forward it to the insulin pump to start the new schedule.  
\begin{algorithm}[htbp]
	\DontPrintSemicolon 
	\KwIn{Prescription control command ($C$)}
	\KwOut{$C$ + Hash($c$)}
	\For{$i \gets 0$ \textbf{to} $n$} {
		$H_c = Hash(C_{i})$\\
		$A_c = Append(C_{i},H_c)$\\
		$S = Init(SSH)$\\
		$Send(A_c,S)$\;
		}
	\caption{Send prescription control command}
	\label{algo:send} 
\end{algorithm}

\begin{figure*}
    \centering
    \includegraphics[width=\textwidth]{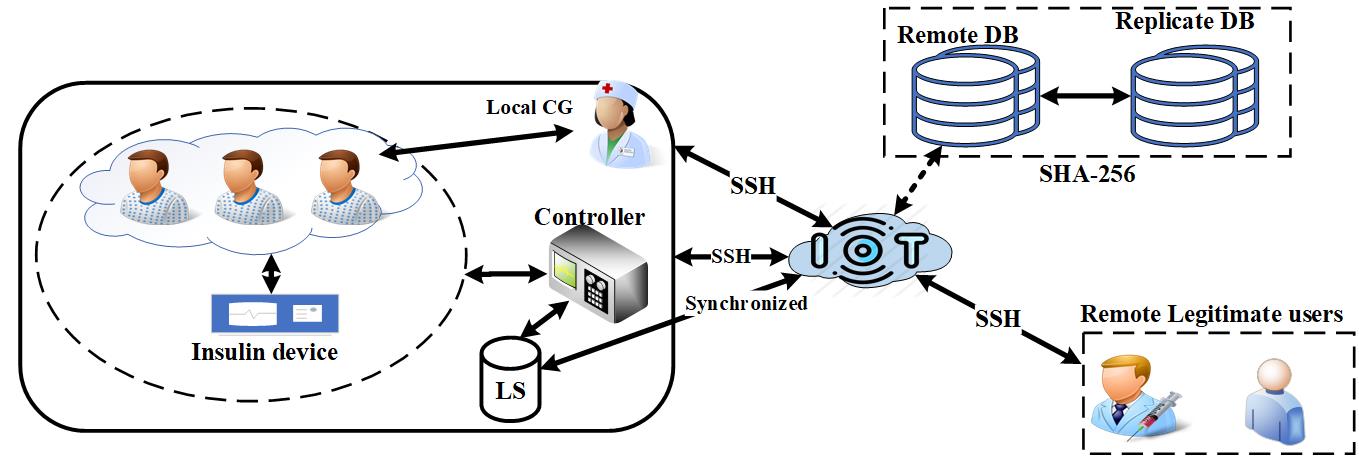}
    \caption{General architecture of the proposed scheme.}
    \label{fig:connection_architecture}
\end{figure*}

Figure~\ref{fig:connection_architecture} shows the connection between different components of IoT health care system. The embedded micro-controller controls the insulin device and collects the required health information. This is accomplished using a serial connection (6.25Mbps) between the micro-controller and the infusion pump. The Cortex-M3 micro-controller, which is embedded in the LPC1768 board, uses a universal asynchronous receiver-transmitter (UART) that supports 8 bits communication without parity and is fixed at one stop bit per configuration. The Keil LPC1768 board is programmed using micro-vision-5 software development kit (SDK) under windows 10 and implemented under \textit{C} software stack.  

The micro-controller collects data and stores them on local storage (LS) and remote storage (Remote DB) through the SSH connection. The IoT-cloud takes the responsibly to provide a replica for the stored data, it is considered as one of the great benefits of using the IoT-cloud. The data between the IoT-cloud and local storage are synchronized all the time to provide quick local access for the \mbox{patient\textquotesingle s} health records. 

The insulin device receives the doses schedule and delivers insulin to the diabetic patient. A local caregiver (CG) is responsible for a group of patients in emergency situations. A patient using the Alaris 8100 infusion pump will take preset insulin doses regularly~\cite{grant2014infusion}. 
The Alaris infusion pump is controlled and monitored by the Keil Cortex M3 board through a serial connection. 
All dosages related records are sent to the cloud through the Keil board using the Ethernet connection. 
To ensure the security and authenticity, the recorded data are digitally signed using the \textit{SHA-256} compression function and encrypted using a symmetric key encryption mechanism. Moreover, the 
The signature and \mbox{patient\textquotesingle s} records are stored together in the cloud.

In the cloud, a Secure Socket Shell (SSH) is provided to authorized entities to access the health records. 
For instance, a physician can follow up with a \mbox{patient\textquotesingle s} case using a mobile application or a web browser. 
Furthermore, research institutions are given the authorization to access health records upon agreements made between
patient, medical centers, and research institutions.

The integrity of the health care records is verified using the SHA-256 signature. While the authenticity is ensured by the encryption mechanism and \textit{SSH} connection. 
The SHA-256 value is computed after the health records or prescription commands are generated. 
Then the generated SHA-256 is appended to the corresponding data (health record or preset control command). 
The health record and its signature remain correlated in all places (cloud, hospital, and \mbox{patient\textquotesingle s} side). 
For instance, the physician in the hospital confirms that the record is received without altering using the SHA-256 signature. 
When the health record is received at the hospital, SHA-256 computation will be carried out. 
The resultant SHA-256 value will be compared with the appended SHA-256 value. 
Once both values are equal, the record will be confirmed to their corresponding patient. 
Otherwise, the health record will be discarded as it does not belong to the patient. Bearing in mind that all connections and data transfer are carried out using an encrypted \textit{SSH} connection. 

In the case of the preset control command, this command is generated from the hospital and appended with its corresponding hash value.
The preset control command and the SHA signature are sent through the cloud to the infusion pump. At the \mbox{patient\textquotesingle s} side, the hardware takes the responsibility to check the genuineness of the received control command by SHA-256 computation and comparison. The Keil micro-controller computes the SHA-256 value for the received preset control command and then compares the result with the appended SHA-256 value. Once authorized, the preset control command is passed to the infusion pump for a new schedule. 

In the case of a fault exception, all Cortex-$M$ processors (including Keil LPC-1768) have a fault exception mechanism embedded inside the processor. If any fault is detected, the corresponding exception handler will be executed~\cite{alkim2016newhope}.

\subsection{Case Study: A Markov Model of proposed scheme}
\begin{figure}[t]
    \centering
    \includegraphics[width=\columnwidth]{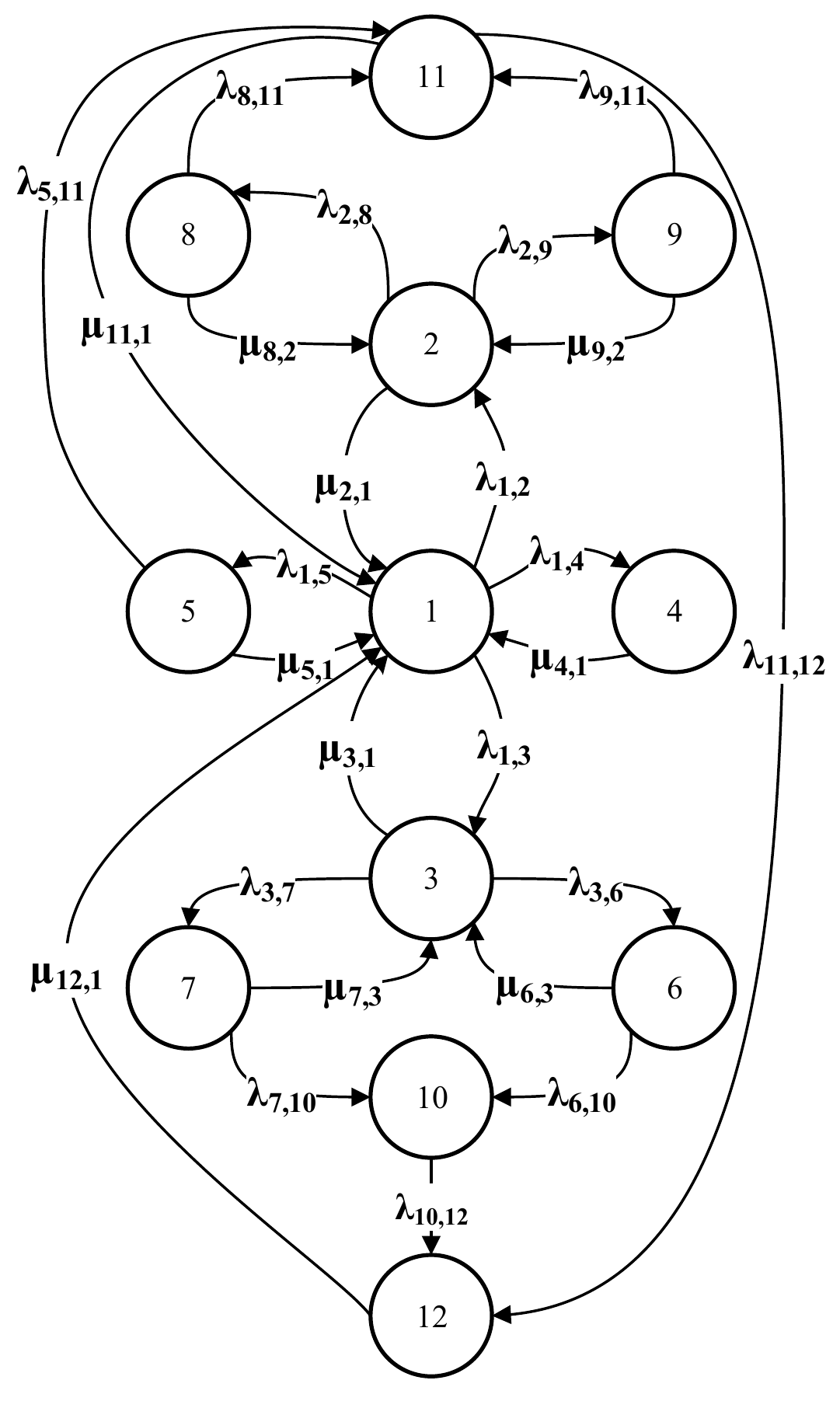}
    \caption{Markov model graph for the IoT health care failure.}
    \label{fig:markov_failure}
\end{figure}

In IoT health care system, the failure of one or more components may lead to system failure. In our design, we have four main components: 
\begin{enumerate*}
    \item Insulin Pump. It is represented by the Alaris 8100 infusion pump. 
    \item Micro-controller. It is represented by the LPC-1768 Keil board.  
    \item IoT-cloud. It provides infrastructure and medium. 
    \item Authority failure that represents the loss of security. 
\end{enumerate*}
Figure~\ref{fig:markov_failure} shows the Markov model that connects the main components during system failure. The failure rate is represented by the symbol $\lambda$ and the recovery rate is represented by the symbol $\mu$. 

The case study depicts 12 states that represent the transition from one state to another with the corresponding failure rate and recovery rate. However, some states are represented by the failure rate only because they are unable to recover. Thereby, the states are defined as follows:
\begin{enumerate*}
    \item Normal operation where all components work as required.
    \item Insulin pump failure due to hardware defects.
    \item IoT-cloud failure due to connection failure.
    \item Failure due to data delivery between Insulin Pump and micro-controller.
    \item Failure due to the power supply.
    \item IoT-cloud software failure.
    \item IoT-cloud hardware failure.
    \item Insulin pump software failure.
    \item insulin pump hardware failure. 
    \item IoT-cloud failure due to the failure of cloud components.
    \item Insulin pump failure due to the failure of insulin pump components. 
    \item Failure of the system.
\end{enumerate*}

The Markov model depicted in Figure~\ref{fig:markov_failure} can be represented as a system of Kolmogrov differential equations, as shown by equations~(\ref{eq:dp1})-(\ref{eq:dp12}). 
The probability ($P_i(t)$) represents the probability to find the system in state $i$. In our design, we chosen the initial conditions as follows: $P_1(t) = 1$, $P_i(t) = 0$ for $i$ = 2, .., 12. 

To collect the failure components and build our case study, we analyzed references ~\cite{paul2011review}\cite{farooq2015critical}\cite{kodeswaran2016idea}\cite{solaiman2016monitoring}\cite{hassanalieragh2015health}\cite{abawajy2017federated}\cite{guenego2016insulin}. All kind of failures are caused by software or hardware failures that might affect the main system components and cause the system failure. To further help other researchers, We list the values of failure and recovery rates in Table~\ref{tab:parameters}. 

\begin{equation}
\begin{split}
     dP_{1}/dt = -(\lambda_{1,2} + \lambda_{1,3} + \lambda_{1,4} + \lambda_{1,5})P_1(t) +\\
      \mu_{2,1}P_2(t) + \mu_{3,1}P_3(t) + \mu_{4,1}P_4(t) + \mu_{5,1}P_5(t)\\
      +\mu_{11,1}P_{11}(t) +\mu_{12,1}P_{12}(t)\\
\end{split}
\label{eq:dp1}
\end{equation}
\begin{equation}
\begin{split}
  dP_2/dt = -(\mu_{2,1} + \lambda_{2,9} + \lambda_{2,8})P_2(t) +\\
  \lambda_{1,2}P_1(t) + \mu_{9,2}P_9(t) + \mu_{8,2}P_8(t) 
   \end{split}
\label{eq:dp2}
\end{equation}
\begin{equation}
\begin{split}
   dP_3/dt = -(\mu_{3,1} + \lambda_{3,6} + \lambda_{3,7})P_3(t)+ \\
 \lambda_{1,3}P_1(t) + \mu_{6,3}P_6(t) + \mu_{7,3}P_7(t) 
   \end{split}
\label{eq:dp3}
\end{equation}
\begin{equation}
\begin{split}
     dP_4/dt = -\mu_{4,1}P_4(t) + \lambda_{1,4}P_1(t)
\end{split}
\label{eq:dp4}
\end{equation}
\begin{equation}
\begin{split}
     dP_5/dt = -(\mu_{5,1}+\lambda_{5,11}) P_5(t) +\lambda_{1,5}P_1(t)
\end{split}
\label{eq:dp5}
\end{equation}
\begin{equation}
\begin{split}
     dP_6/dt = -(\mu_{6,3}+\lambda_{6,10}) P_6(t) +\lambda_{3,6}P_3(t)
\end{split}
\label{eq:dp6}
\end{equation}
\begin{equation}
\begin{split}
     dP_7/dt = -(\mu_{7,3}+\lambda_{7,10}) P_7(t) +\lambda_{3,7}P_3(t)
\end{split}
\label{eq:dp7}
\end{equation}
\begin{equation}
\begin{split}
     dP_8/dt = -(\lambda_{8,11} + \mu_{8,2}) P_8(t) +\lambda_{2,8}P_2(t)
\end{split}
\label{eq:dp8}
\end{equation}
\begin{equation}
\begin{split}
     dP_9/dt = -(\mu_{9,2}+\lambda_{9,11}) P_9(t) +\lambda_{2,9}P_2(t)
\end{split}
\label{eq:dp9}
\end{equation}
\begin{equation}
\begin{split}
     dP_{10}/dt = -\lambda_{10,12}P_{10}(t) +\lambda_{6,10}P_6(t) + \lambda_{7,10}P_7(t)
\end{split}
\label{eq:dp10}
\end{equation}
\begin{equation}
\begin{split}
    dP_{11}/dt = -(\mu{11,1}+\lambda_{11,12}) P_{11}(t) +\lambda_{9,11}P_9(t) + \\
     \lambda_{8,11}P_8(t) + \lambda_{5,11}P_5(t)
\end{split}
\label{eq:dp11}
\end{equation}
\begin{equation}
\begin{split}
     dP_{12}/dt = -\mu_{12,1}P_{12}(t)+\lambda_{10,12}P_{10}(t) \\+ \lambda_{11,12} P_{11}(t)
\end{split}
\label{eq:dp12}
\end{equation}

In the subsequent section, we show the performance characteristics and their applicability to our proposal. 
\section{Results and Discussion}
\label{sec:section5}
Our proposal has been tested toward the five performance characteristics that are mentioned in Section~\ref{sec:section2}. 

\subsection{Availability}
As mentioned earlier, the availability property ensures that the system is available all the time. Our design is tested for availability by solving the system of Kolmogorov differential equations and compute the probabilities of system states. The values of system sates probabilities, after calculations, are as follows:
\begin{align*}
\centering
    & P_1 = 0.9925712 & P_2 = 0.0002091\\
    & P_3 = 0.0005966 & P_4 = 0.002998966\\
    & P_5 = 0.00009805 &P_6 = 1.09{E-}06 \\
    & P_7 = 2.99{E-}05 & P_8 = 0.0019989 \\
    & P_9 = 0.00049866 & P_{10} = 4.24{E-}07\\
    & P_{11} = 0.0009958 & P_{12} = 3.00{E-}07
\end{align*}

The availability function is represented by the probability value of $P_1(t)$, which means that the system has a probability of $\approx 99.26 \%$ to stay at the normal state. The calculated probability proves the availability property of the IoT health care embedded scheme. Through this value, the proposed design ensures a high level of availability.    

\subsection{Confidentiality}   
To provide a confident system for data on transit, our design uses the \textit{SSH} tunnel that is only given to the authorized entities. The \textit{SSH} connection is initialized only by a legitimate user and supported by "private-public key pair authentication" scheme that ensures the connection is established between the designated two parties. 

\subsection{Integrity}
\begin{figure}[htbp]
\centering
        \includegraphics[scale=0.75]{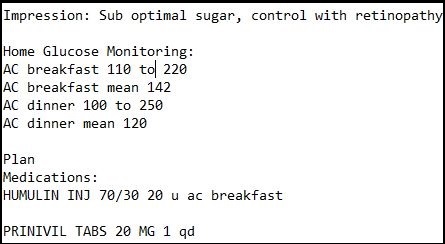} 
        \caption{Snipped health record from the original sample.}
                \label{fig:sample_orig}
\end{figure}

\begin{figure}[htbp]
\centering
        \includegraphics[scale=0.75]{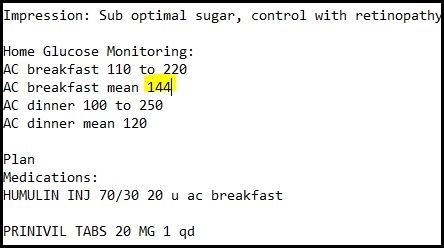} 
        \caption{Snipped health record from the modified sample.}
                \label{fig:sample_mod}
\end{figure}

The proposed design has been tested and verified for integrity using sample data from~\cite{BibEntry2018Oct}. 
The sample data contains glucose levels in the \mbox{patient\textquotesingle s} body during a 24 hour period, a \mbox{patient\textquotesingle s} profile information, and the \mbox{patient\textquotesingle s} medical information. A snipped portion of the sample data is shown in Figure~\ref{fig:sample_orig}, the figure shows the glucose levels in the \mbox{patient\textquotesingle s} body after two meals (breakfast and dinner). To test the integrity property, the sample data is modified as shown in Figure~\ref{fig:sample_mod}. When both figures are compared, the only difference between them is the "AC breakfast Mean", it is equal to 142 in the original sample and 144 in the modified one.


The proposed design considers that the SHA-256 value is computed every time a health record is requested. The sample data is stored in the cloud and appended with the corresponding SHA-256 value. If the \mbox{patient\textquotesingle s} side requests the same health record, the micro-controller will compute the SHA-256 value of the record and compares it with the appended SHA-256 value. If both hash values (cloud and patient) are equal then the received record is valid and never been tampered during the transmission. Table~\ref{tab:sample} shows the SHA-256 value of the sample record on both sides where the sample record has not tampered.  

However, any tiny modification to the health record will produce a totally different SHA-256 hash value. Table~\ref{tab:sample_modified} shows two different hash values for the original sample that is requested from the cloud side and the modified sample at the \mbox{patient\textquotesingle s} side. Both SHA-256 values are different because the received record on the \mbox{patient\textquotesingle s} side has been altered during transmission. Then, the micro-controller at the receiver side will detect the alteration after comparing both hash values. 

\begin{table}[htbp]
    \centering
    \caption{SHA-256 HASH VALUES OF THE SAMPLE DATA ON BOTH SIDES.}
            \begin{tabular}{ll}
        \textbf{Cloud side}:& \begin{tabular}[c]{@{}l@{}}\texttt{14b93acf-ccdcbe40-ea3795be-c1073498-}\\
                                                                                        \texttt{51a96c90-6cedfc9c-49d8e2cf-a141befb} \\ 
                    \end{tabular}  \\ \hline
        \textbf{Patient side}:& \begin{tabular}[c]{@{}l@{}}\texttt{14b93acf-ccdcbe40-ea3795be-c1073498-}\\
                                                                                            \texttt{51a96c90-6cedfc9c-49d8e2cf-a141befb} 
                \end{tabular} 
        \\ \hline
    \end{tabular}
    \label{tab:sample}
\end{table}

\begin{table}[htbp]
    \centering
        \caption{SHA-256 HASH VALUES OF THE ORIGINAL AND MODIFIED SAMPLE DATA ON BOTH SIDES.}
            \begin{tabular}{ll}
        \textbf{Cloud side}:& \begin{tabular}[c]{@{}l@{}}\texttt{14b93acf-ccdcbe40-ea3795be-c1073498-}\\
                                                                                        \texttt{51a96c90-6cedfc9c-49d8e2cf-a141befb} \\ 
                    \end{tabular}  \\ \hline
        \textbf{Patient side}:& \begin{tabular}[c]{@{}l@{}}\texttt{358c4f29-f0e2bb60-8efa35d4-a88a6b3b-}\\
                                                                                            \texttt{58939ffd-deebf824-8065c195-b834b8cd} 
                \end{tabular} 
        \\ \hline
    \end{tabular}
    \label{tab:sample_modified}
\end{table}

On another hand, to ensure the integrity of prescription control command, the same procedure is carried out between the sender (corresponding physician) and receiver (micro-controller). At the \mbox{patient\textquotesingle s} side, the micro-controller detects the alteration and discard the tampered control commands.

\subsection{Authentication}
To provide an authentic system, the SSH protocol is employed to ensure that only legitimate users are eligible to access the health records. Moreover, in the case of the prescription control command, special users are given a special SSH tunnel and a public-private key pair to ensure the security and authenticity of the communication medium between the Caregiver (CG) and micro-controller. 

\subsection{Authorization}
The authorization and verification of certain actions before execution are accomplished by the encrypted \textit{SSH} connection and the \textit{SHA}, respectively. The encryption of health records ensures that only the authorized entities can decrypt and read the data contents. Moreover, if any certain action is tampered or modified before reaching the destination, then the corresponding hash value will determine whether the action is authorized. Moreover, the patient is given some privileges to change the schedule according to a predefined prescription from the corresponding physician. 

\subsection{Speed}
\begin{figure*}
    \centering
    \includegraphics[angle=270,width=\textwidth]{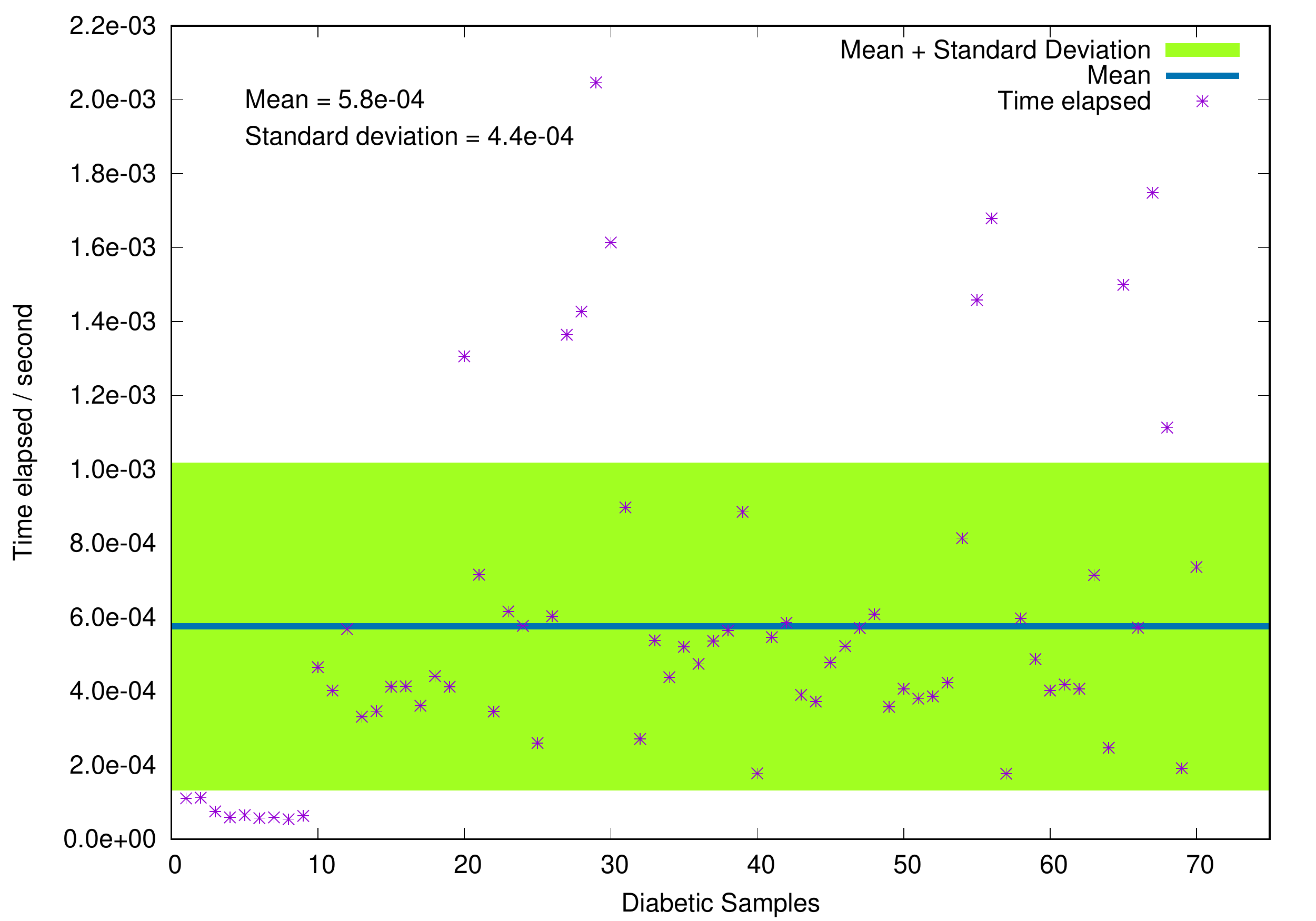}
    \caption{Time elapsed to process 70 diabetic samples.}
    \label{fig:time_elapsed}
\end{figure*}

The processing speed of the proposed design is tested using 70 samples of diabetic's records~\cite{BibEntry2019Feb}. Figure~\ref{fig:time_elapsed} shows the time elapsed (in second), mean and standard deviation of the 70 samples. The elapsed time to process the samples depends on different factors, including, sample size, connection speed, and system utilization. The figure shows how the processing speed changes according to the aforesaid factors. The average time to process these samples is equal to $5.8{e}-04$-second, while the standard deviation value shows the amount of variation of the elapsed time for all samples. 

\subsection{Final Remarks}
Our design provides a set of benefits to the health care systems, particularly, diabetic patients. We list these benefits as follows:
\begin{itemize}
    \item Patients can access their health records easily and communicate with their caregiver instantly. 
    \item Caregivers and physicians can control the insulin infusion pump remotely according to reliable information delivered through the proposed design. 
    \item The security and integrity of patient's records are guaranteed by the encrypted \textit{SSH} and the \text{SHA}. 
\end{itemize}
However, the limitation of this approach can be seen in the case of a successful attack on the used security components. Until now, there is no successful collision attack for the \textit{SHA-256} that is used in this design. The collision attack allows an adversary to tamper the data contents and produce the same hash (signature) of data before and after modification. Moreover, the length extension attack is a kind of attacks that targets the keyed hash algorithms (HMAC). Therefore, we avoid using the (HMAC) in our design and keep the authenticity requirements to the symmetric encryption mechanism of the \textit{SSH} connection.       

\section{Conclusion and Future work}
\label{sec:section6}
In this paper, a reliable embedded health care system based on the Internet of Thing is presented. The proposed design employs secure hash algorithm SHA-256, Secure Socket Shell (SSH), Keil LPC-1768 board, Alaris 8100 infusion pump, and IoT-cloud to build the health care system. The proposed design showed that the reliability characteristics of availability, confidentiality, integrity, authentication, and authorization are accomplished. Moreover, the results showed that the proposed design has a $99.3\%$ probability to stay in the normal operation stage and an average speed of $5.8\times10^{-04}$ seconds to process the health records.   

The scope of reliable IoT-based health care system is open. In the future, further analysis of the health care system to develop a generalized reliability model of the health care system including handheld medical devices.

\section*{Acknowledgments}
This publication was funded by a grant from the United States Government and the generous support of the American people through the United States Department of State and the United States Agency for International Development (USAID) under the Pakistan - U.S. Science \& Technology Cooperation Program. The contents do not necessarily reflect the views of the United States Government. 

Computing services, financial and administrative support from the North Dakota State University Center for Computationally Assisted Science and Technology (CCAST) and the Department of Energy through Grant No. DE-SC0001717 are gratefully acknowledged.   
\appendix
The values of failure and recovery rates, which were used in the case study, are listed in Table~\ref{tab:parameters}.
\begin{table}[ht!]
\label{tab:parameters}
\caption{FAILURE AND RECOVERY RATES PARAMETERS}
\begin{centering}
\begin{tabular}{cccc}
\toprule 
Failure ($\lambda$) & Value & Recovery ($\mu$) & Value\tabularnewline
\midrule
\midrule 
$\lambda_{1,2}$ & 1.857E-09 & $\mu_{2,1}$ & 99.57E-2\tabularnewline
\midrule 
$\lambda_{1,3}$ & 2.499E-07 & $\mu_{3,1}$ & 95.08E-2\tabularnewline
\midrule 
$\lambda_{1,4}$ & 3.331E-07 & $\mu_{4,1}$ & 98.76E-2\tabularnewline
\midrule 
$\lambda_{1,5}$ & 4.985E-07 & $\mu_{5,1}$ & 92.37E-2\tabularnewline
\midrule 
$\lambda_{2,8}$ & 2.50E-07 & $\mu_{6,3}$ & 2.12E-3\tabularnewline
\midrule 
$\lambda_{2,9}$ & 2.50E-07 & $\mu_{7,3}$ & 4.07E-3\tabularnewline
\midrule 
$\lambda_{3,6}$ & 7.50E-3 & $\mu_{8,2}$ & 4.20E-4\tabularnewline
\midrule 
$\lambda_{3,7}$ & 3.56E-05 & $\mu_{9,2}$ & 2.93E-4\tabularnewline
\midrule 
$\lambda_{6,10}$ & 1.28E-2 & $\mu_{11,1}$ & 1.23E-6\tabularnewline
\midrule 
$\lambda_{7,10}$ & 1.63E-2 & $\mu_{12,1}$ & 1.857E-8\tabularnewline
\midrule 
$\lambda_{8,11}$ & 2.00E-4 &  & \tabularnewline
\midrule 
$\lambda_{9,11}$ & 3.11E-5 &  & \tabularnewline
\midrule 
$\lambda_{10,12}$ & 2.70E-3 &  & \tabularnewline
\midrule 
$\lambda_{11,12}$ & 25.87E-3 &  & \tabularnewline
\bottomrule
\end{tabular}
\par\end{centering}
\label{tab:parameters}
\end{table}

\bibliographystyle{IEEEtran}
\bibliography{bibo.tex}
\end{document}